\documentclass[structabstract]{aa}
\usepackage{graphicx}
\usepackage{txfonts}
\usepackage{natbib}
\def\simless{\mathbin{\lower 3pt\hbox
     {$\rlap{\raise 5pt\hbox{$\char'074$}}\mathchar"7218$}}}   
\def\simmore{\mathbin{\lower 3pt\hbox
     {$\rlap{\raise 5pt\hbox{$\char'076$}}\mathchar"7218$}}}   

\def\msun{~{\rm M}_\odot}

\begin{document}
\title{
Inclination effects on the X-ray emission of Galactic black-hole binaries
}

\subtitle{}

\author{
Pablo Reig\inst{1,2}
\and
Nikolaos D. Kylafis\inst{2,1} 
}

\institute{
Institute of Astrophysics, Foundation for Research and Technology-Hellas, 71110 Heraklion, Crete, Greece
\and
University of Crete, Physics Department \& Institute of
Theoretical \& Computational Physics, 70013 Heraklion, Crete, Greece\\
}

\date {Received ; Accepted ;}


\abstract
{
Galactic black-hole X-ray binaries (BHBs)
emit a compact, optically thick, mildly
relativistic radio jet when they are in the hard and hard-intermediate states.
In these states, BHBs exhibit a correlation 
between the time lag of hard photons with
respect to softer ones and the photon index of the power law component that
characterizes the X-ray spectral continuum above $\sim$ 10 keV. The correlation,
however, shows large scatter.
In recent years, several works have brought to light the importance of taking into
account the inclination of the systems to understand  the X-ray and radio
phenomenology of black-hole binaries.
}
{
Our objective is to investigate the role that the inclination
plays on the correlation between the time lag  and the photon index.
}
{
We have obtained {{\it RXTE}} energy spectra and light curves of a sample of 
black-hole binaries with different inclination angles. We have computed  the
photon index and the time lag between hard and soft photons and 
have performed a
correlation and linear regression analysis of the two variables.
We have also computed energy spectra and light curves of black-hole binaries
using the Monte Carlo technique that reproduces the process of 
Comptonization in the jet.
We account for the inclination effects by recording the photons that escape from 
the jet at different angles. From the simulated 
light curves and spectra we have obtained model-dependent photon index and time lags that 
we have compared with those obtained from the real data.
}
{
We find that the correlation between the time lag and 
the photon index is tight in
low-inclination systems and becomes weaker in high-inclination systems. The
amplitude  of the lags is also larger at low and intermediate inclination angles
than at high inclination. We also find that the photon index and the time lag,
obtained from the simulated spectra and light curves, also follow different
relationships for different 
inclination angle ranges. Our jet model reproduces the
observations remarkably well. The same set of models that reproduces the
correlation for the low-inclination systems, also  accounts for  the correlation for
intermediate- and high-inclination systems fairly well.
}
{
The large dispersion observed in the time lag - photon index correlation in
BHBs can naturally be explained as an inclination effect.
Comptonization in the jet explains the steeper dependence of the lags on the
photon index in low/intermediate-inclination systems than in 
high-inclination ones. 
}


\authorrunning{}

\titlerunning{Inclination effects on BHBs }

\maketitle

\begin{table*}
\centering
\caption{List of outbursts and sources. The fourth column gives the total exposure time of all the
observations analyzed. The fifth column gives the number of observations used in the 
spectral and timing analysis (blue empty circles in Fig.~\ref{hid}) over the total number  
(blue empty circles plus black dots in Fig.~\ref{hid}). 
}
\label{sources}
\begin{tabular}{llccclc} 
\hline
Object 		&Outburst	&MJD interval	 &Total exposure &Number of	&Inclination   &$N_{\rm H}$ \\
		&epoch		&		 &time (ks)      &observations	&	 &$\times 10^{22}$ cm$^{-2}$ 	        \\
\hline
4U\,1543--475	&2002		&52443--52565	&275.4		&20/112		&Low		&0.43 (1)       \\
MAXI\,J1836--194&2011		&55804--55896	&116.9		&69/74		&Low		&0.3 (7)        \\
Cyg X--1	&2003--2004	&52693--53182	&367.4		&147/151	&Low		&0.54 (3)       \\
\hline
GX\,339--4	&2002		&52311--52884	&495.6		&48/267		&Intermediate 	&0.4 (1)	 \\
 		&2004		&53050--53498	&565.2		&152/328 	&		 \\
 		&2007		&53769--54678	&572.6		&191/347	&		 \\
 		&2010		&54889--55618	&326.0		&107/317    	&		 \\  
XTE\,J1650--500	&2002		&52159--52447	&327.9		&47/182 	&Intermediate 	&0.7 (1)	 \\
Swift\,J1753.5-0127&2005	&53553--53819	&188.1		&72/73		&Intermediate	&0.2 (5)	 \\
		&2007--2011	&54226--55894	&607.2		&275/278	&		 \\
\hline
XTE\,J1550--564	&2000		&51644--51741	&128.7		&36/236		&High	  	&0.65 (1)	\\
		&2001--2004	&51938--53163	&306.5   	&69/101		&		\\
GRO\,J1655--40	&2005		&53423--53685	&2238.3		&68/501		&High	 	&0.8 (1)	\\
H\,1743--322	&2003		&52740--53055	&668.0		&42/194		&High		&2.4 (1)		\\
		&2004		&53197--53329	&79.1		&19/49		&		&		\\
		&2005		&53595--53668	&46.4		&11/23		&		&		\\
		&2008a		&54746--54789	&90.8		&18/27		&	   	&	        \\
		&2008b		&54743--54849	&68.6		&36/38		&	   	&	        \\
		&2009		&54980--55039	&93.4		&15/51		&	   	&	        \\   
		&2010a		&55220--55245	&47.8		&21/33		&		&		\\
		&2010b		&55418--55470	&106.8		&25/58		&		&		\\
		&2011		&55664--55735	&48.2		&19/39		&	   	&	        \\
\hline
XTE\,J1752--223	&2009		&55130--55414	&408.2		&113/207	&Intermediate	&0.6 (6)	 \\
XTE\,J1817--330 &2006		&53768--53950	&382.3		&20/140		&High?    	&0.15 (1)	\\	
GS\,1354--645	&1997		&50774--50840	&50.5		&7/9		&High		&3.72 (1)	\\
MAXI\,J1543-564	&2010		&55465--55567	&154.0		&6/100		&High		&1.4 (2)	\\
Swift\,J1842.5--1124&2008	&54656--54803	&83.2		&31/49		&High		&0.4 (8)	\\
XTE\,J1859+226  &1999		&51463--51749	&356.5		&25/127		&High    	&0.34 (1) 	\\
XTE\,J1118+480	&2000		&51633--51764   &138.7		&47/50		&High    	&0.01 (1)	\\ 
		&2005		&53383--53428   &71.7     	&27/39		&		&		\\
MAXI\,J1659--152&2010		&55465--55567	&140.9		&39/66		&High?		&0.27 (4)	\\
\hline
\multicolumn{7}{l}{(1) \citet{dunn10}; (2) \citet{stiele12}; (3) \citet{hanke09}; (4) \citet{jonker12}}\\
\multicolumn{7}{l}{(5) \citet{cadolle07}; (6) \citet{chun13}; (7) \citet{russell14}} \\
\end{tabular}
\end{table*}

\begin{figure*}
\begin{center}
       \includegraphics[width=14cm]{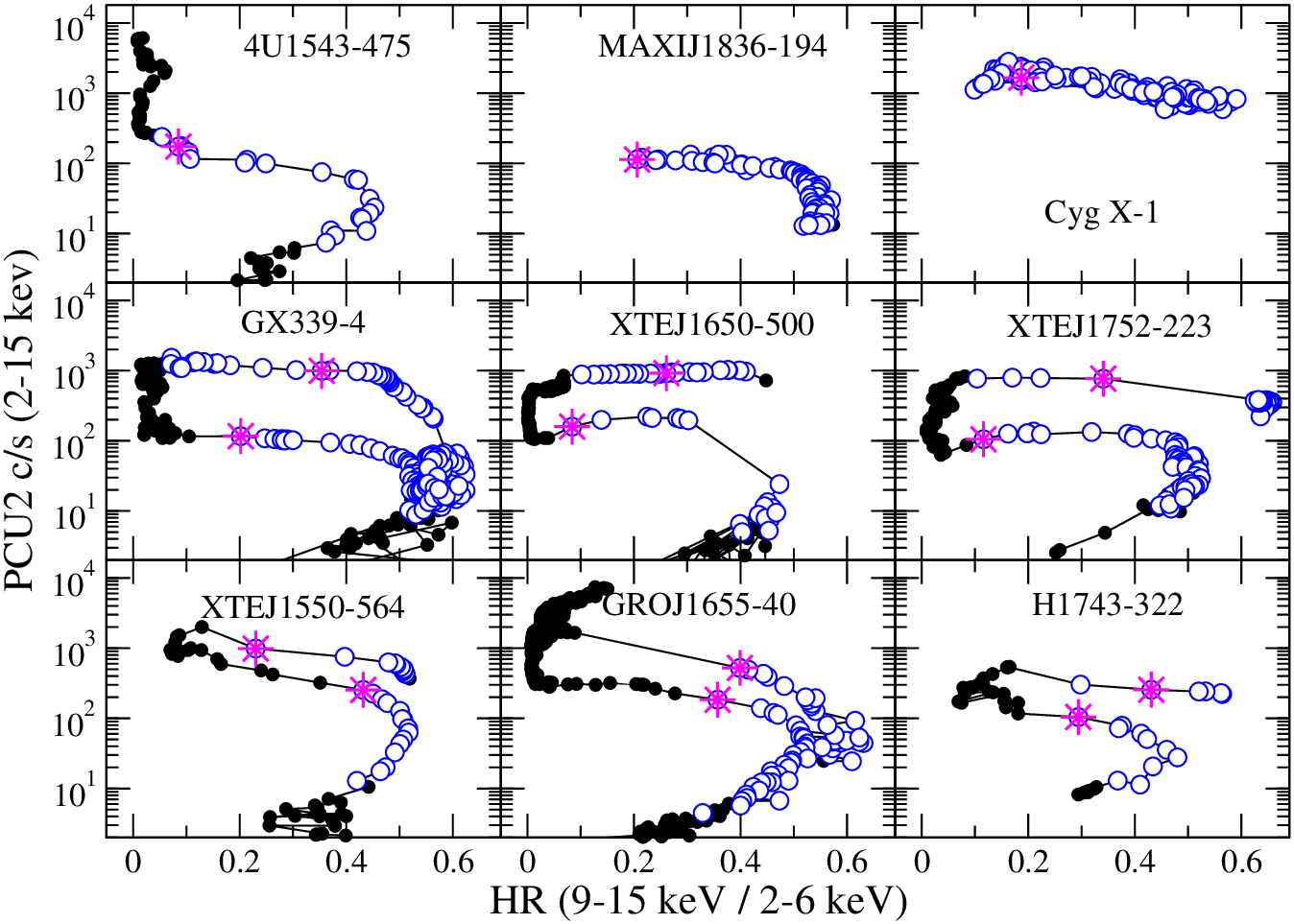}
\end{center}
    \caption{Hardness-intensity diagrams. Each point corresponds to one
    observation. The blue empty circles identify the observations used
    in the final lag-spectral analysis. The magenta stars indicate observations
    with a photon index close to $\Gamma \approx 2$, which roughly 
separates the HS from the HIMS.}
    \label{hid}
\end{figure*}

\section{Introduction}

Observations over more than 30 years have revealed a common phenomenology in the
X-ray properties of Galactic black-hole binaries (BHB). Most of them are
transient sources that become active during a short period of time (several
months) compared to the time they are in an off state (years). The realization
that the source goes through different states during outbursts represents an
important step forward in our understanding of the observational properties of
BHBs. Each state is characterized by certain spectral (e.g., slope of the X-ray
spectral continuum) and temporal (e.g., shape of the power spectral density,
frequency of quasi-periodic oscillations) parameters, which vary smoothly within
a given state, but may show sudden changes when the source changes state. An
effective way to separate the observations into states is to plot the
hardness - intensity diagram (HID). A BHB traces a
$q$-shaped curve as the outburst progresses
\citep{homan05,remillard06,belloni10}. Roughly speaking, the left and
right branches correspond to the soft and the hard state, 
respectively, while the lines that
connect these branches correspond to the intermediate state. The fact that the
horizontal lines do not overlap, but are traced at different X-ray luminosity,
indicates a strong hysteresis effect. An explanation of the hysteresis pattern
and the direction in which the source moves in the HID (anticlockwise) has been
given by \citet{kylafis15}.

When BHBs are in the hard (HS) and hard-intermediate (HIMS) states, their X-ray
spectrum in the 2-200 keV band is well represented by a power-law function that
falls exponentially at high energies. Reflection components, such as iron line
emission at 6.4-6.6 keV and an excess of emission around 20-30 keV, are commonly
observed. BHBs also exhibit strong emission in the radio band, whose origin is
attributed to a compact, partially optically thick, mildly relativistic jet.

There is growing evidence that the inclination of the binary orbit with respect
to the observer's line of sight plays an important role in the determination of
the characteristics of the detected emission. \citet{ponti12} studied the winds
emitted by the accretion disk and concluded that, because of the small opening
angles, they are only observed  in high-inclination systems.
\citet{munoz-darias13} showed that inclination has a strong effect on the
evolution of BHBs through the HID. The $q$-track in the HID of low-inclination
systems displays more square shape, while that of high-inclination BHBs has
more triangular shape. They also found that the accretion disks in
high-inclination systems look hotter than in low-inclination systems.

The results of these two works are based on observations mainly in the soft
state, when the radio jet is absent. Inclination-dependent differences have also
been reported in the HS and the
HIMS, when the jet is present. \citet{heil15} found that
high-inclination BHBs display larger hardness ratios than low-inclination
systems with similar power spectral shape. The rms variability, however, was not
seen to be different in low- and high-inclination systems.

\citet{motta15} found that the amplitude of low-frequency type-C quasi-periodic
oscillations (QPOs) is stronger for nearly edge-on systems (high inclination),
while type-B QPOs are stronger when the accretion disk is closer to face-on (low
inclination). In contrast, the broad-band noise associated with type-C QPOs is
stronger in low-inclination sources. They concluded that these two types of QPOs
and the broad-band noise associated with type-C QPOs correspond to different
phenomena. While type-C QPOs are consistent with the truncated disk model
and arise from relativistic precession
of the inner hot flow, type-B QPOs are likely associated with the
radio jet. The broad-band noise likely comes from fluctuations in the mass
accretion rate.

\citet{vandereijnden17} performed a systematic analysis of the
inclination dependence of phase lags associated with both Type-B and Type-C QPOs
in Galactic black hole binaries. They found that the phase lag at the
Type-C QPO frequency strongly depends on inclination, both in evolution with QPO
frequency and sign. As the QPO frequency increases, the low-inclination systems
tend to display larger positive (i.e. hard) lags, while high-inclination systems
turn to negative (i.e. soft) lags.

Finally, \citet{motta18} investigated the correlation between the radio and the X-ray
emission in BHBs and found that high-inclination objects tend to be radio-quiet,
while low-inclination systems appear to be radio-loud.

Recently, we performed a detailed study of the HS and 
the HIMS of BHBs as a class
and found a correlation between the photon index of the power-law component and
the time lag of the hard photons with respect to the softer ones
\citep{reig18,kylafis18}. We showed that up-scattering of low-energy photons
(from the accretion disk) by highly-energetic electrons (in the jet) can explain
the correlation. Although the correlation is statistically significant, it
exhibits a large amount of scatter.  The main goal of the present work is to
investigate the effect of the orbital inclination on the correlation between the
time lag and the photon index in BHBs. As in previous works that study the
effects of inclination on the properties of BHBs, we assume that there is no
intrinsic physical difference among BHBs, hence any systematic difference in the
correlation at different inclinations must be attributed to this parameter. We
conclude that the scatter in the correlation 
found by Reig et al. (2018) can be explained as an inclination
effect. 
{\it The scatter is due to the fact that we see the systems at different jet
viewing angles.}

\section{Observations and data analysis}

The data presented in this work has been downloaded from the Rossi X-ray Timing
Explorer ({\it RXTE}) archive. We have followed the same analysis procedure as
explained in \citet{reig18}, with the only difference that here we have used the
proportional counter array (PCA) only, to minimize the effect of reflection on
the X-ray continuum. 

Due to RXTE's low-Earth orbit, the observations consist of a number of
contiguous data intervals or {\em pointings} (typically 0.5--1 h long)
interspersed with observational gaps produced by Earth occultations of the
source and passages of the satellite through the South Atlantic Anomaly. For
each observation, we have obtained the average energy spectrum  over the
energy range 2--25 keV and the light curves in the energy ranges $2-6$ keV,
$9-15$ keV, and $2-15$ keV. The number of observations is given in
Table~\ref{sources}. Due to the varying number of detectors (from 1 to 5) in the
observations and to avoid calibration effects, for the spectral analysis,
including the HID, we used only the proportional counter unit 2 (PCU2). For the
timing analysis, we have used all PCUs that were on during the observations,
because we have performed the timing analysis on segments of data 64 seconds
long. In this way we increase the signal-to-noise with respect to the case of
one detector only. To have a reasonable quality of the observations, we have
considered only those with  an average count rate\footnote{The reason that the
rate of some data points (blue circles) goes below 20 c s$^{-1}$ in
Fig.~\ref{hid} is because that figure was made using rates from one PCU only,
while the data selection was done with all the PCUs that were on.} of at least
20 c s$^{-1}$ in the 2--15 keV energy range and at least 640 s of contiguous
observations, i.e. without gaps (10 segments of 64 s each).  To  extract the
observations that correspond to the HS and HIMS, we have selected observations
with rms, in the 0.01--30 Hz frequency range and in the  2-15 keV energy range,
larger than 10\%. The time resolution of the light curves used to create
the power spectra and to perform the time-lag analysis was $2^{-7}$ s.

The time lag has been computed for the $9-15$ keV photons  with respect to $2-6$
keV ones and has resulted from  the average of the time lag in the frequency
range 0.05--5 Hz. To obtain the photon index, we  have fitted an absorbed broken
power-law model to the spectral continuum. The hydrogen column density  has been
fixed to the values given in Table~\ref{sources}. A narrow  Gaussian component
(line width $\sigma \simless 0.9$ keV)  has been added to account for the iron
line emission at around 6.4 keV. The photon index used in our analysis is the
one that corresponds to the hard power law, that is, after the break.  In
support of the use of this phenomenological model is the fact that the reduced
$\chi^2$ in 94\% of the fits resulted in a value smaller or equal than 2 (85\%
smaller than 1.5).

\subsection{Source selection}

As a starting point, we have used the list of sources presented in Table 1 of
\citet{motta18}. However, because the inclination effects may be rather subtle,
we wish to have the cleanest possible sample of sources.  Thus, we have selected
sources with {\em i)}  well-sampled outbursts or a large number of observations
in the {\it RXTE} archive, {\em ii)}  densely populated HS, and {\em iii)}  
well-constrained values of the inclination (see \citealt{motta15} and
\citealt{motta18} for an explanation of the method employed to infer the system
inclination). We have chosen systems for which the range of
the inclination values, given in \citet{motta18},  clearly fall in one of the
following categories: low-inclination BHBs (Li-BHBs), intermediate-inclination
BHBs (IMi-BHBs), and high inclination BHBs (Hi-BHBs) depending on whether the
angle between the observer's line of sight and the perpendicular to the orbital
plane is smaller than 35$^{\circ}$, between 35$^{\circ}$ and 70$^{\circ}$, and
larger than 70$^{\circ}$, respectively. 

These constraints reduced the number of sources to nine, three for
each group. These are (see Table~\ref{sources}): 4U\,1543--475,
MAXI\,J1836--194, and Cyg X--1 (Li-BHBs),  GX\,339--4, XTE\,J1650--500, and
Swift\,J1753.5--0127 (IMi-BHBs) and  XTE\,J1550--564, GRO\,J1655--40, and
H\,1743--322 (Hi-BHBs). We shall discuss other sources in Sect~\ref{other}.

\section{Results}

We begin our analysis by generating the HID for each source as shown in
Fig.~\ref{hid}. In this Figure, each data point represents the average
count rate of one observation and was obtained using 16-s binned light curves.
The black filled circles represent  the complete data set, while the blue empty
circles are the observations that we have selected to obtain time lag and
photon index. They correspond to the HS and the HIMS. Star symbols separate
approximately the HS from the HIMS and roughly corresponds to a photon index of
$\Gamma\approx2$ (see below). 

Fig.~\ref{indiv} shows the correlation between the time lag and the photon
index for Li-BHBs  (top panel), IMi-BHBs (middle panel), and Hi-BHBs (bottom
panel) for individual systems and individual outbursts. Li-BHBs show a very
tight correlation. As the inclination increases, the scatter increases. 
To produce Fig.~\ref{indiv}, the individual observations of each source were
binned in  $\Gamma$ bins of size $\Delta\Gamma=0.1$. The data points correspond
to the weighted average of all the observations that fell in the corresponding
bin. 

Fig.~\ref{aver} shows the average behavior for each one of the three groups in
Fig.~\ref{indiv}. Fig.~\ref{aver} has been generated in the same way as
Fig.~\ref{indiv}, but this time all the observations of all the sources of the
same group and of the same $\Gamma$ bin have been
merged together and averaged.

\begin{figure}
\centering
\includegraphics[angle=0,width=9cm]{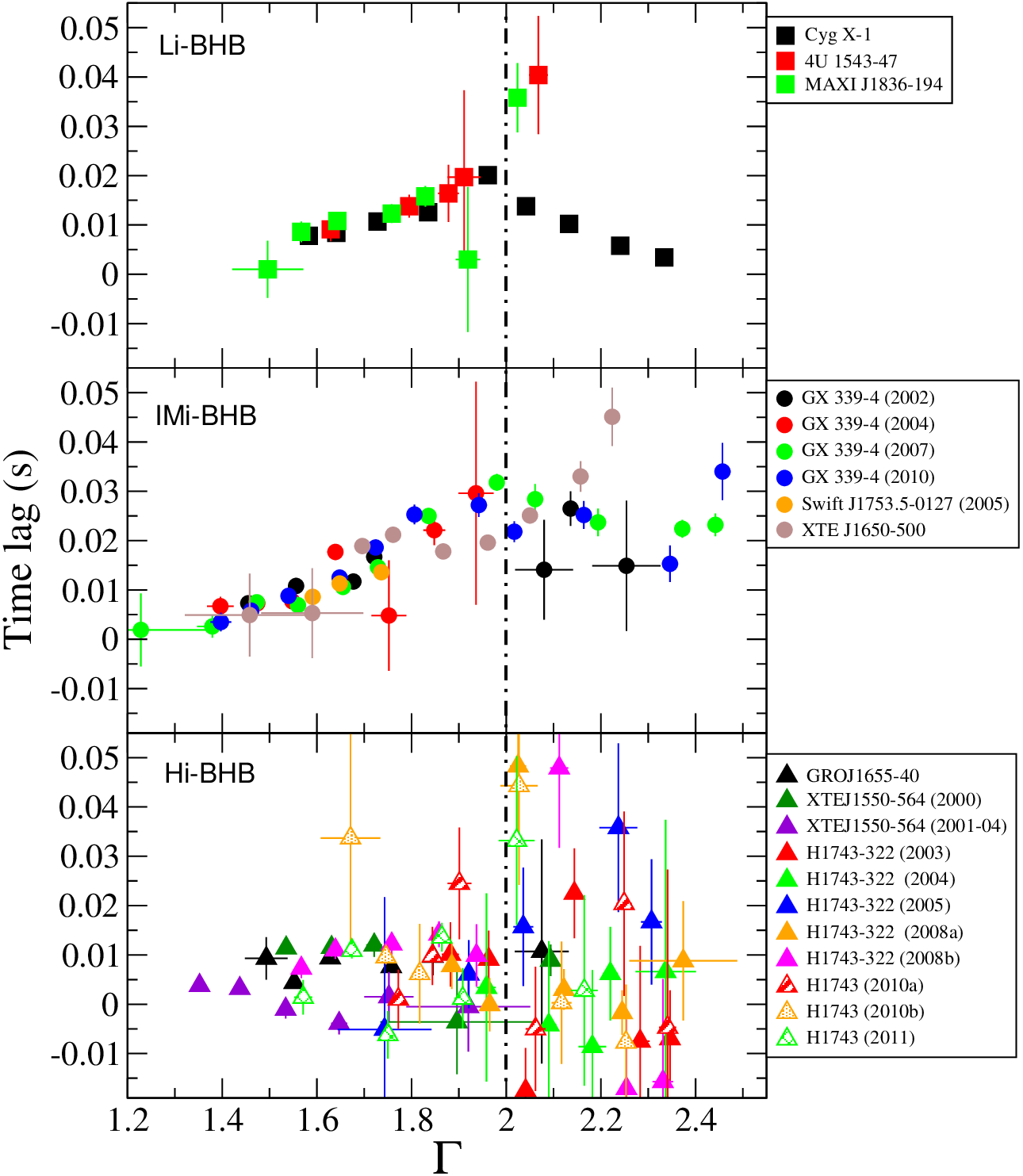} 
\caption{Correlation between the time lag and the photon index for individual
systems. {\em Top panel:} low-inclination systems. {\em Middle panel:}
intermediate-inclination systems. {\em Bottom panel:} high-inclination systems.
}
\label{indiv}
\end{figure}

\begin{figure}
\centering
\includegraphics[angle=0,width=8cm]{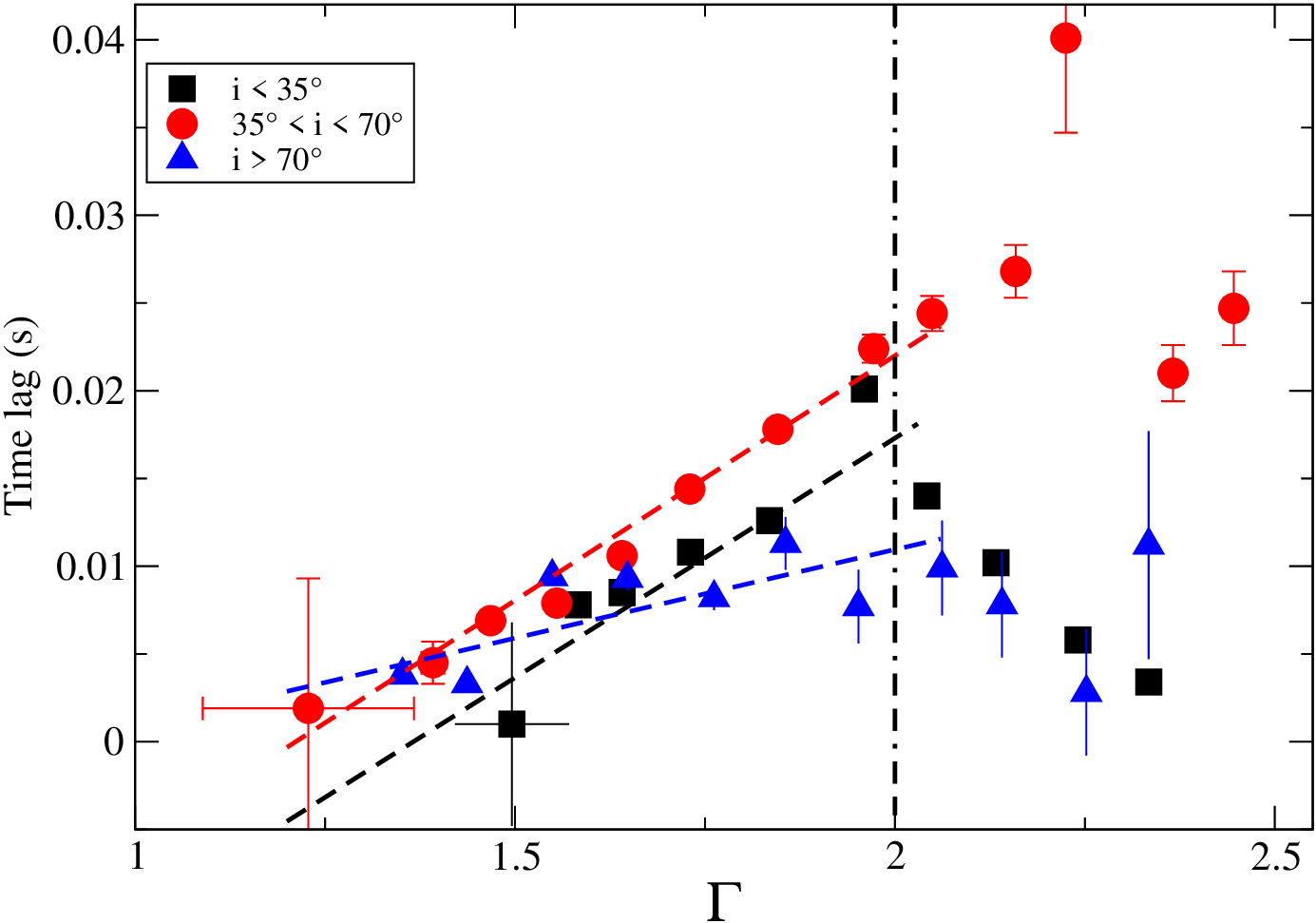} 
\caption{Average correlation between the time lag and the photon index for
Li-BHBs  ($i \le 35^{\circ}$, black squares), 
IMi-BHBs ($35^{\circ} < i \le 70^{\circ}$, red dots), 
and Hi-BHBs ($i > 70^{\circ}$, blue triangles).}
\label{aver}
\end{figure}

\subsection{Correlation analysis}

We have performed three different, but related, analyses.  First, we have
examined how strong the correlation between the time lag and the photon index
is.  This has been done with the correlation coefficient $\rho$. The closer
$\rho$ is to 1, the stronger the correlation is.  Second, we have checked
whether the correlation is statistically significant.  This has been done by
testing the null hypothesis that the two variables are uncorrelated, using
the $t$-statistic  $t=\rho\sqrt{(N-2)/(1-\rho^2)}$, where $N$ is the number of
data points.  The smaller the
probability $p$, the more significant the correlation is.  Finally, we have
performed a linear regression analysis to fit a straight line to the data.  We
have used the bisector BCES (bivariate  correlated errors and intrinsic scatter)
method  following \citet{akritas96}. This method  takes into account both the
individual errors and the intrinsic scatter. Moreover, this method is more
appropriate than the traditional least squares estimator, because, since there
is no a priory reason to choose one of the two variables as the independent
variable, the BCES method uses the bisector of the two lines that correspond to
the least-square fit of Y on X and X on Y.

The precise instant at which the source transits from the HS to the HIMS is
difficult to identify with X-ray data only, because the properties of the HIMS
are consistent with being the extension of those of the HS \citep{belloni10}.
The $t_{\rm lag}-\Gamma$ correlation is positive at low $\Gamma$ and
becomes negative at high $\Gamma$.  Although the point at which the slope of the
$t_{\rm lag}-\Gamma$ correlation changes may not coincide with the transition
from the HS to the HIMS,  the correlation becomes negative during most part of
the HIMS \citep{kylafis18}.  We have  performed our analysis on the data points
that contribute to the positive correlation, that is, for $\Gamma \leq 2$. In
terms of spectral states,  our analysis includes all the HS and perhaps the
onset of the HIMS for some sources.  To facilitate the visualization, we have
drawn a vertical dashed-dotted line in Figs.~\ref{indiv} and \ref{aver} and
marked  the points with $\Gamma\approx2$  with a magenta star in Fig.~\ref{hid}.

\begin{table*}
\centering
\caption{Results of the linear regression and correlation analysis  for 
$\Gamma \le 2$.  Here $\rho$ is Pearson's correlation coefficient, 
$N$ is the number of points, and $p$ the probability that the null hypothesis
(no correlation) is true. "Low" refers to systems with $i \le 35^{\circ}$,
"Intermediate" to $35^{\circ} < i \le 70^{\circ}$, "High" to $i > 70^{\circ}$.}
\label{tab1}
\begin{tabular}{lccccc} 
\hline
\hline
		&Slope		&Intercept		&$\rho$ &N	&$p$-value	        \\
\hline
Low	   	&$0.0273\pm0.0061$ &$-0.0373\pm0.0101$  &0.88	&7	&$8.4\times10^{-3}$    	   \\
Intermediate	&$0.0279\pm0.0024$ &$-0.0338\pm0.0041$  &0.99	&9	&$2.4\times10^{-7}$    	   \\
High	  	&$0.0101\pm0.0030$ &$-0.0093\pm0.0049$  &0.75	&8	&$5.0\times10^{-2}$   	   \\
\hline
\end{tabular}
\end{table*}

We have a pair $(\Gamma,t_{\rm lag})$ for each observation. Because of the large
number of observations analyzed and for the sake of clarity,  as we discussed
above, the data presented in Figures 2 and 3 have been binned in  $\Gamma$ with
a bin size of   $\Delta \Gamma = 0.1$.  The results of the analysis are given in
Table~\ref{tab1} for the data sets shown in Fig.~\ref{aver} and can be
summarized as follows:

\begin{itemize}

\item Li-BHBs and IMi-BHBs display a distinct and strong
correlation with correlation coefficients $\rho \simmore 0.9$. In Hi-BHBs, the
correlation is weaker.

\item The correlation is significant above 99\% confidence level for Li-BHB and
IMi-BHBs and $\simmore$95\% for Hi-BHBs. In other words, the probability that
the two variables are uncorrelated is small.

\item As the inclination increases, the scatter of the correlation also
increases (Fig.~\ref{indiv}). 

\item At $\Gamma \simless 1.6$, the amplitude of the lags is similar for all
systems. Above this value, IMi-BHBs show, on average, longer time lags
than the rest.

\item The slope of the linear regression of the Li-BHBs is consistent with that
of the IMi-BHBs within errors and significantly ($4.5\sigma$ and
$11.5\sigma$, respectively) different from zero. In Hi-BHBs, the slope is
different from zero at $\sim 3\sigma$.

\end{itemize}

\begin{figure}
\centering
\includegraphics[angle=0,width=8cm]{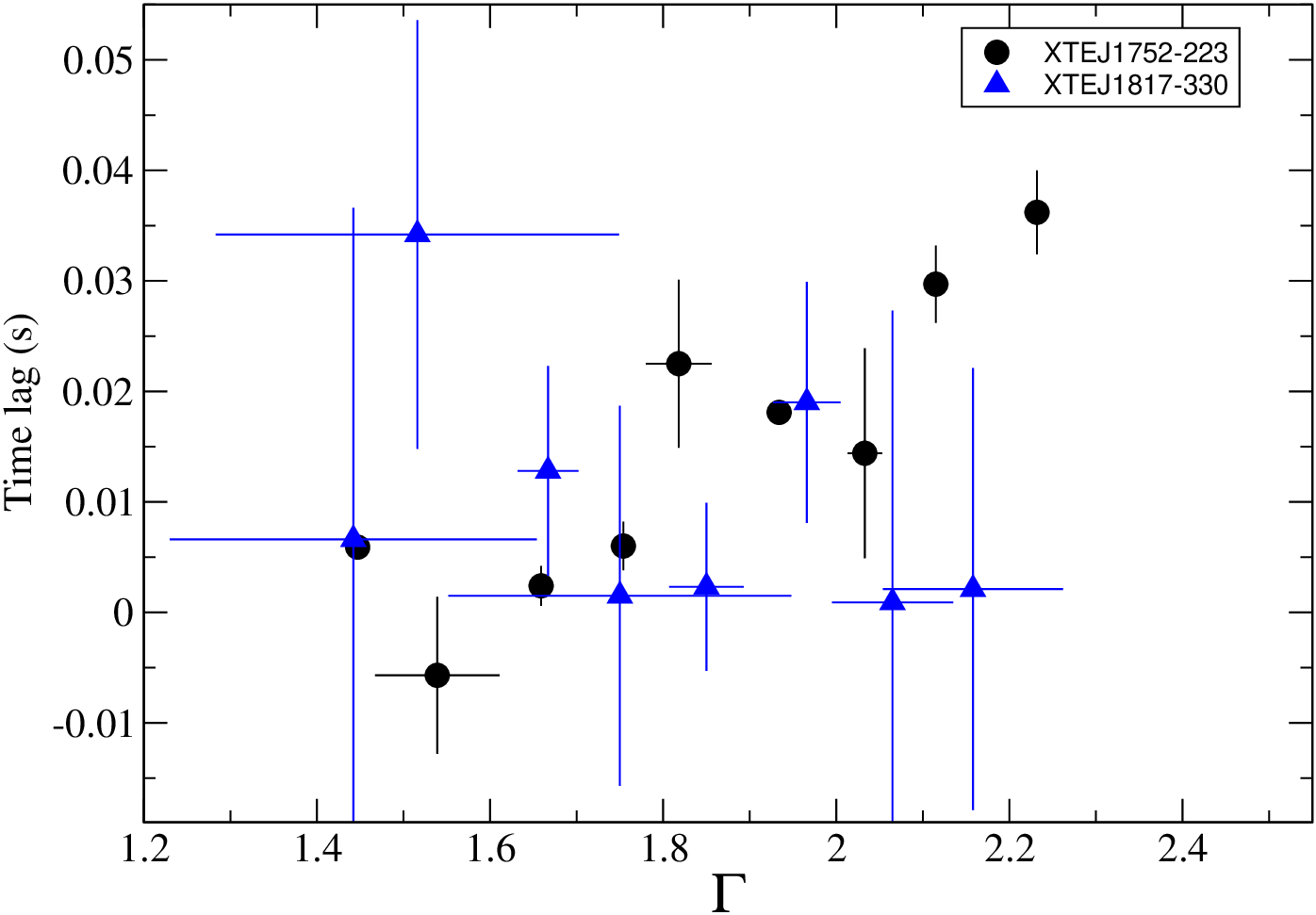} 
\caption{Time lag versus photon index for
XTE\,J1752-223 and XTE\,J1817--330.}
\label{unknown}
\end{figure}
\begin{figure}
\centering
\includegraphics[angle=0,width=8cm]{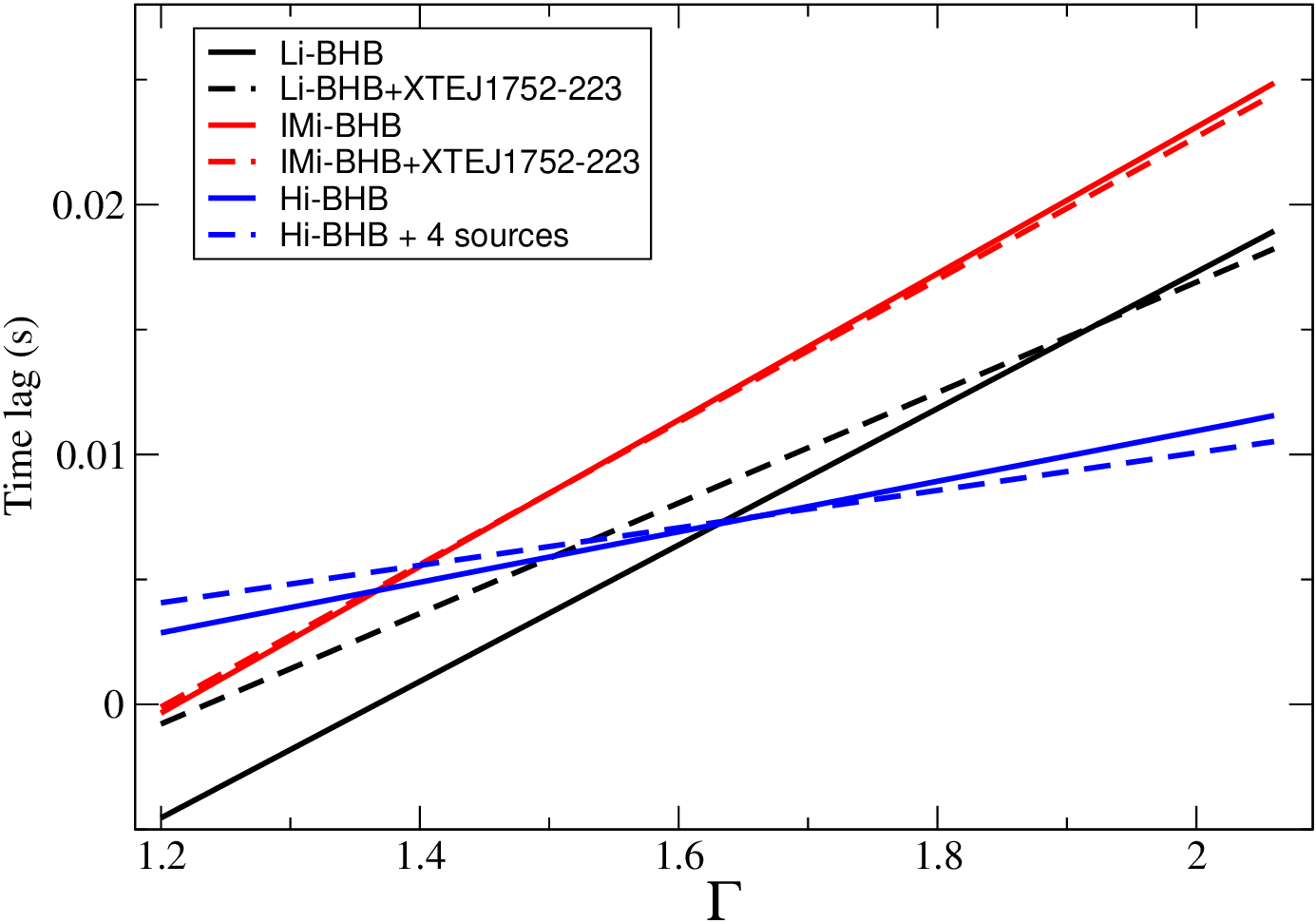} 
\caption{Average correlation between the time lag and the photon index for
different selections of sources. The blue dashed line corresponds to the Hi-BHB
group with the addition of GS\,1354--645, XTE\,J1859+226, Swift\,J1842.5--1124, 
and MAXI\,J1543-564.  
}
\label{slopes}
\end{figure}

\subsection{Other sources}
\label{other}

In this section, we investigate how the selection of sources  affects the
statistical analysis performed above. In Table~1 of \citet{motta18}, there are
two sources (XTE\,J1752-223 and XTE\,J1817--330) with an estimated inclination
of $5-60^{\circ}$. Thus they belong to either the Li-BHB  or the IMi-BHB
group. In addition, another three sources  have intermediate inclination,
but we do not include these three sources in our analysis. 
A\,0620--00 due to lack of $RXTE$ observations, 
GRS\,1915+105 due to its peculiar
behavior, and MAXI\,J1659--152 because its inclination is uncertain: while
\citet{motta18} give an inclination angle in the rage $30-70^{\circ}$ based on
the overall amplitude of the type-C QPO, \citet{kuulkers13} suggest $i \sim
65-80^{\circ}$ based on the presence of intensity drops, possibly attributed to
absorption dips or partial eclipse.

Figure~\ref{unknown} shows the $\Gamma - t_{\rm lag}$ pairs for 
XTE\,J1752-223 and XTE\,J1817--330. While XTE\,J1752-223 shows
a clear correlation, confirming it as a Li-BHB or IMi-BHB, the relationship
between the lags and the photon index in XTE\,J1817--330 resembles that of
Hi-BHBs. As \citet{motta15} pointed out, the inclination measurements mainly
rely on the assumption that absorption dips and wind-related features are a
strong indication of high-orbital inclination. No such features have been
detected in XTE\,J1817--330. However, this non-detection does not guarantee that
they may not appear in future observations. 
Thus,  a Li-BHB, whose classification is
based on these criteria, may turn into a Hi-BHB if new observations reveal dips
in the X-ray light curve. The opposite, i.e., a Hi-BHB turning into a Li-BHB,
cannot happen.

In addition to XTE\,J1550--564, GRO\,J1655--40, and H\,1743--322, the list of
sources of \citet{motta18} contains other Hi-BHBs with an apparently reliable
estimate of the inclination: XTE\,J1908+094, XTE\,J1118+480, GS\,1354--645,
V404\,Cygni, and XTE\,J1859+226.  Of these five sources, we have not analyzed
XTE\,J1908+094, because during most of the observations it was offset by $>
25\arcmin$ with respect to the center of the field of view, since there was a
nearby strong X-ray source 4U\,1907+09 (that was actually the target of the
observation). We remind that the PCA/{\it RXTE} had no imaging capabilities and
that the field of view was $\sim 1^{\circ}$. The {\it RXTE} archive does not
have data of V404\,Cygni. We have also ignored XTE\,J1118+480 because the
observations sample a very narrow part of the HS branch. In fact, during the
2005 outburst, the observation seems to cover the transition from the HS to the
quiescent state at the very end of the decaying phase of the outburst.  
Another five sources, XTE\,J1748-288, Swift\,J1842.5--1124, IGR\,J17177--3656,
4U1630--47, and MAXI\,J1543-564 are classified as high-inclination objects,
without specifying any value for the inclination angle.  XTE\,J1748-288  and
IGR\,J17177--3656 suffer from source confusion as there are other sources in the
field of view and are offset with respect to its center.  Also, 4U1630--47 was
always in the soft state.  The inclusion of GS\,1354--645,  XTE\,J1859+226,
Swift\,J1842.5--1124, and MAXI\,J1543-564 to the initial group of Hi-BHBs does
not alter  significantly the slope of the correlation. 

Figure~\ref{slopes} shows the slopes for different selections of sources. We do not
see substantial differences. The largest difference is found when the source
XTE\,J1752-223 is assumed to be a Li-BHB. However, no difference in the slope of
the correlation is found if the source is included in the IM-BHB group.
{\it Therefore, we conclude that the inclination angle of XTE\,J1752-223 must be
in the range $35^{\circ}-70^{\circ}$. }

\subsection{Jet model: Monte Carlo simulations}

We have developed a model that simulates the process of inverse
Compton scattering in a jet. 
Low-energy photons, presumably from the accretion disk, are
up-scattered by energetic electrons moving outwards at 
mildly relativistic speeds in
the jet. We assume the jet to be parabolic with a finite acceleration region. 

The parameters of the model are:  the optical depth along the jet's axis
$\tau_{\parallel}$, the width of the jet at its base $R_0$,  the parallel, 
$v_0$, and perpendicular, $v_{\perp}$, components of the velocity, or
equivalently the Lorenzt factor $\gamma=1/\sqrt{1-(v_0^2+v_{\perp}^2)/c^2}$, the
distance $z_0$ of the bottom of the jet from the black hole, the total height
$H$ of the jet, the temperature  $T_{bb}$ of the soft-photon input, the size
$z_1$ and the exponent $p$ of the acceleration zone,
where $v_{\parallel}(z) =(z/z_1)^p ~
v_0$, for $z \le z_1$,  and $v_{\parallel}(z)=v_0$ for $z>z_1$.

The jet model used here is the same as the one used in \citet{reig18} and
\citet{kylafis18}. The novelty of the version used in this work is that we
now compute the dependence of the escaping photons as a function of
the angle $\theta$ between the observer and the jet
axis. We have used ten different bins in $w=\cos\theta$, 
each with size $\Delta w=0.1$, covering the range
from $w=0$ to $w=1$. In terms of $w$, the different categories of sources are
defined as: $0.8 < w \le 1$ for Li-BHBs, $0.3 < w \le 0.8$ for
IMi-BHBs, and $0 \le w \le 0.3$ for Hi-BHBs. 
We note that we have not found
substantial differences by considering $0.9 < w \le 1$ for Li-BHBs and $0.3
< w \le 0.9$ for IMi-BHBs. Because the number of photons that scatter in the
direction perpendicular to the jet axis ($0 \le w \le 0.3$) is
considerably smaller than at other directions, we increased the number of input
photons by an order of magnitude with respect to our previous works. Here
we have used $10^8$ photons. 
This increases the computing time but ensures good statistics in all the bins. 
Typically, the
fraction of photons that escape from the jet distributes as 2--3\% in
$w=0.0-0.3$, 35--45\% in $w=0.3-0.8$, and 40-45\% in $w=0.8-1.0$. The rest,
about $10-20$\%, are emitted toward the accretion disk.

The various Monte Carlo models that we have run differ only in 
the value of the optical depth $\tau_{\parallel}$ (or equivalntly in the
density) and the width of the jet at its base
$R_0$, while the rest of the parameters
are fixed  at the following reference values:  $z_0 = 5 r_g$, $H=10^5 r_g$,
$v_0=0.8 c$, $v_{\perp}=0.4 c$, $z_1= 50 r_g$, $p=0.5$, and $T_{bb} = 0.2$ keV,
where  
$r_g=GM/c^2$ is the gravitational radius. We assume a black-hole mass of 10
$\msun$.

For each model and each angle bin, the code generates an energy spectrum  and
two light curves in the same energy bands as those chosen 
for the data analysis, namely $2-6$ keV and $9-15$ keV.
The relevant parameters in this
study are the photon index of the power-law spectral continuum and the time lag
between the detection of hard photons 
($9-15$ keV) with respect to softer photons ($2-6$ keV). 

The shape of the simulated energy spectra is indeed well represented by a
power-law with a roll over at high energies, as observed in real data.  To
determine the model photon index, we fit the spectra with a power-law 
and an exponential cutoff.  The light curves are processed in the same way as
the real light curves to derive the time lag. We compute the cross-spectrum,
defined as $C(\nu_j)=X_1^*(\nu_j)X_2(\nu_j)$, where 
$X_i(\nu_j)$ is the Fourier transform of the time series
and the asterisk denotes complex
conjugate. The phase lag between the signals in the two bands at Fourier 
frequency $\nu_j$ is $\phi(\nu_j)=\arg[C(\nu_j)]$ [the position angle of
$C(\nu_j)$ in the complex plane]  and the corresponding time lag $t_{\rm
lag}(\nu_j)= \phi(\nu_j)/2\pi\,\nu_j$.

\begin{figure}
\centering
\includegraphics[angle=0,width=8cm]{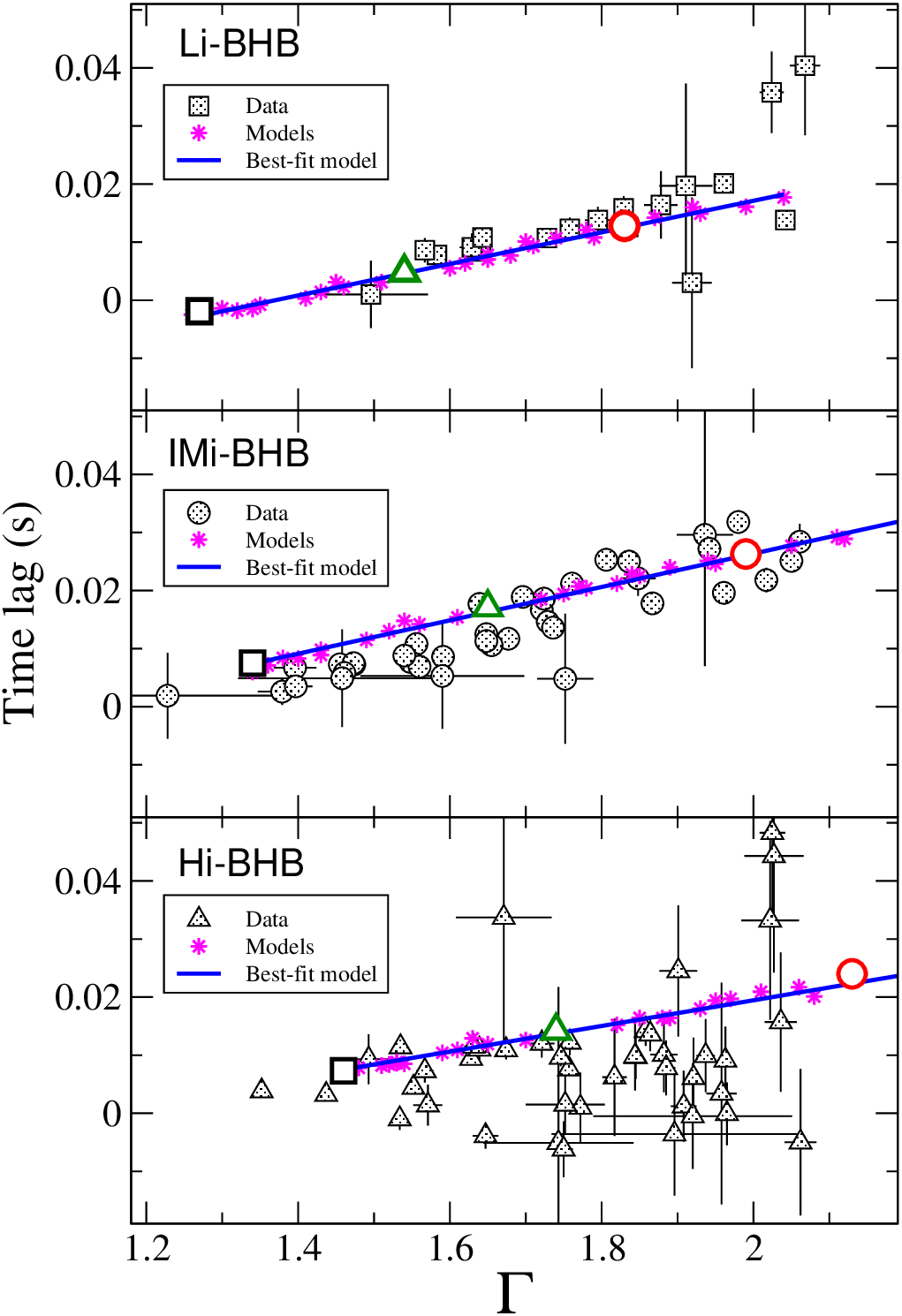} 
\caption{Comparison of data (dot-filled symbols) and models (magenta stars). The
lines represent the best linear fit to the models. The three larger symbols
correspond to three representative models: 
$\tau_{\parallel}=10$ and $R_0= 50r_g$ (square),
one with $\tau_{\parallel}=5$ and $R_0= 140r_g$ (triangle), and
one with $\tau_{\parallel}=2.75$ and $R_0= 250r_g$ (circle).}
\label{bestmod}
\end{figure}

\section{Comparison of the model with the observations}

Our aim is to investigate whether Comptonization in an extended jet can explain
the different correlations between the photon index and the time lag shown in
Fig~\ref{indiv}.

We assume that the jet axis is perpendicular to the orbital plane, hence the
orbital inclination and the observation
angle $\theta$ coincide. In other words, if $w \sim
1$, the observer sees the jet along its axis, whereas  $w \sim 0$ corresponds
to systems in which the observer sees the jet perpendicularly.

We have proceeded as follows: 
first we have chosen the models that reproduce the linear fit
of Li-BHBs (top panel in Fig.~\ref{indiv} and black line and black squares in
Fig.~\ref{aver}). That is, by changing $\tau_{\parallel}$ and $R_0$, 
we have built a
series of models, whose $\Gamma_{\rm Li-BHB}$ and $t_{\rm lag,Li-BHB}$ match
those of the observations. These two values have been obtained using 
only data that
fell in the $w=0.8-1$ range, hence the subindex. Then, using {\em the same
models}, i.e., the same pairs  $(\tau_{\parallel}, ~R_0)$, we have
obtained ($\Gamma_{\rm IMi-BHB}$, $t_{\rm lag,IMi-BHB}$) and ($\Gamma_{\rm
Hi-BHB}$, $t_{\rm lag,Hi-BHB}$), which now correspond to $w=0.3-0.8$ and
$w=0-0.3$, respectively.

Our results are plotted in Fig.~\ref{bestmod}.  The black symbols in this Figure
are the same as the colored symbols in Fig.~\ref{indiv} (for $\Gamma \simless
2.1$), while the  magenta stars correspond to the models. The line is simply the
best linear fit to the model data points and is plotted for clarity.  Given the
simplicity of the model and the complexity and dissimilarity of the data, the
agreement is remarkable.

To furher understand the effects of inclination, we have selected three models:
one with $\tau_{\parallel}=10$ and $R_0= 50r_g$ (big black empty square in 
Fig.~\ref{bestmod}),
one with $\tau_{\parallel}=5$ and $R_0= 140r_g$ (big green empty triangle in 
Fig.~\ref{bestmod}), and
one with $\tau_{\parallel}=2.75$ and $R_0= 250r_g$ (big red empty circle in 
Fig.~\ref{bestmod}).  
It is evident from the three panels
of Fig.~\ref{bestmod} that {\it the same jet model produces different
spectra and different time lags} at different inclinations.

\citet{heil15}
showed that high-inclination systems {\em with a similar
power spectral shape} tend to show harder emission (measured as a hardness
ratio) than lower inclination systems. Unfortunately, we cannot test this result
with our model, because we cannot identify similar "timing
states". In
fact, our model predicts that the X-ray emission is softer at
high inclination (note the shift toward the right of the three models in each
panel of Fig.~\ref{bestmod}). That is, if we could see the same source from
different viewing angles, we would find a softening of the spectrum (i.e.,
larger photon index) with increasing inclination. 
This does not necessarily mean that high
inclination systems always display softer spectra than low-inclination systems. A
high-inclination system may have a harder spectrum than a low-inclination
system, but if we could see a high-inclination system at a lower angle, then our
model predicts that the X-ray spectrum would be even harder. 
Unfortunately, we cannot see the same source from different viewing angles.

The detection of the 6.4 keV line implies that a reflection component
should contribute to the continuum as well, possibly affecting the power-law
photon index. However, we do not think that our results are significantly
affected by reflection. First, \citet{bagri18}  did not find significant
differences in the photon index after adding reflection in the hard state of
GX\,339--4. They also performed their analysis using {\it RXTE}/PCA data over a
similar energy range to the one we used. Second, we restrict the energy range of
the spectral analysis to $E < 25$ keV, i.e. leaving outside the energies at
which the hump is most prominent. Finally, it appears that reflection dominates
once the source moves well inside the HIMS \citep{plant14}, while we limited
most of our study to the HS.

\section{Discussion}

In a recent work \citep{reig18}, we found that BHBs as a class  exhibit a
correlation between the power-law photon index  and the time lag between hard
and soft photons.  When considered as a whole (many sources), the correlation
shows a large amount of scatter. In this work, we show that the large scatter of the
correlation can be explained as an inclination effect. Low- and
intermediate-inclination systems show a steeper correlation than
high-inclination systems.

In \citet{reig18}, we demonstrated that Comptonization in the jet reproduces
satisfactorily the observed relationship between the X-ray spectral continuum
emission and the time lag of hard photons with respect to softer ones. However,
due to the large dispersion in the data,  concerns about how significantly the
jet model can constrain the correlation remained. After all, fitting a model to
a cloud of points is much less constraining that fitting a model to a tight
correlation. These concerns vanished when we reproduced the very tight $t_{\rm
lag}-\Gamma$ correlation in GX\,339--4 \citep{kylafis18}. In  the present work,
we go one step further in demonstrating the potential of our jet model by
reproducing the observed different $t_{\rm lag}-\Gamma$ correlations of BHBs as
a function of inclination. {\em We emphasize that the same set of models, that
is, the same combination of $\tau_{\parallel}$ and $R_0$  that were selected to
fit the Li-BHBs correlation were also used to reproduce the IMi-BHB and
Hi-BHBs correlations and that the only difference was the angle at which the
escaping photons were recorded}.

Hi-BHBs exhibit a flatter relationship between photon index and lags. The
largest difference is found at larger $\Gamma$,
i.e., at low values of $\tau_{\parallel}$. We explain
this as follows: the low$-\Gamma$ part of the correlation corresponds
to a pure HS, in which presumably the jet is evolving. In
\citet{kylafis18}, we showed that in this state the radius at the base of the
jet increases and the optical depth along the jet decreases as $\Gamma$
increases.  At low $\Gamma$, the jet is narrow and dense
(large $\tau_{\parallel}$). Photons suffer a 
large number of scatterings in the lower part of the jet
(acceleration region),
but with a small mean free path, 
and escape the jet almost in all directions. As the
jet radius increases and the optical depth decreases, photons are able to
travel a larger distance along the jet and enter the region of large
flow speed. Forward scattering of photons by the fast-flowing electrons results 
in their escape at small to moderate angles $\theta$.
Toward the
end of the HS, a fully developed jet is present. Its size is maximum and its
optical depth is moderate. In this configuration, due to the bulk motion of the
electrons in the outflow, most photons are scattered in the direction of 
the flow, have relatively large mean free paths
and travel large distances, thus the hard photons have a long time lag.  
On the other hand,
very few photons escape perpendicular to
the jet axis (Hi-BHBs), and those that do have traveled short
distances. Hence, high-inclination hard photons have short
time lag.

\subsection{Comptonization in the jet: a model that cannot be ignored}

Transient BHBs are excellent laboratories to probe the Physics of accretion and
relativistic ejection of matter. Over the past 20 years, and thanks to the
state-of-the-art space detector technology on board X-ray missions, the
phenomenological description of these states has reached an unprecedented
degree of detail. 
We are now able to monitor changes in the broad (continuum) and
discrete (lines) components of the energy spectrum and of a large number of
timing parameters (rms, QPOs, lags). This extraordinary amount of information
has not translated into a unified physical model and disagreement in even the
most basic level exists. While there is general consensus that hard photons are
produced by inverse Compton scattering, the geometry and properties of the
comptonizing medium are still highly debated. Different models associate the
comptonizing medium with different geometries. It could be an optically thin,
very hot "corona" in the vicinity of the compact object
\citep{titarchuk80,hua95,zdziarski98}, an advection-dominated accretion flow
\citep{narayan94,esin97}, a low angular momentum accretion flow
\citep{ghosh11,garain12}, the base of a radio jet
\citep{band86,georganopoulos02,markoff05}.

Since 2003, we have been demonstrating that Comptonization 
in an extended jet not
only reproduces the general spectral and timing properties of 
transient BHBs, but
also the more stringent constraints imposed by the correlation between spectral
and timing parameters. In addition to quantitatively explaining the emerging
spectrum from radio to hard X-rays \citep{giannios05} and  the evolution of the
photon index and the time (phase)-lags as functions of Fourier frequency
\citep{reig03}, improved versions of our original jet model have been able to
explain the correlation between the time lag and the X-ray photon index in Cyg
X--1 \citep{kylafis08}, in GX339--4 \citep{kylafis18}, 
and in the class of BHB as a
whole \citep{reig18}. 
Similarly for the correlation between time lag and cut off energy in
GX339--4 \citep{reig15}.

In this work, we have shown that the inclination of the system has a profound
effect on the correlation between photon index and time lag. We have  tested
our model against this highly constraining result and passed it with success.
Given the heterogeneity of the data, Fig.~\ref{bestmod} represents 
a huge success of the model.

When one looks at the overall picture, BHBs as a whole show a similar
pattern of variability, supporting the view that the
process of accretion of matter and ejection of a jet is similar in all BHBs.
After all, the definition of source states has been possible thanks to the
repeatability of the behavioral patterns as the source goes through an
outburst.  The reality, however,  
is more complex than the overall picture.  Separate outbursts even
from the same source can look different. The HS may appear straight or slightly
curved in the HID. The loop of the $q$ curve may have a 
rectangular or triangular
shape rather than a rounded shape. A recent study has shown that a substantial
fraction ($\sim$40\%) of the outbursts in transient BHBs do not reach the soft
state \citep{tetarenko16}. Differences between sources are most likely related
to fundamental properties of black holes, such as mass and spin, while
differences between outbursts of the same source can be attributed to different
mass-transfer-accretion rates.

The mass and presumably the spin in transient BHBs vary significantly.
We lack studies that link the actual values of these two parameters with
the physics of accretion and ejection of matter. It is likely that different
combinations of these parameters generate jets that differ in size, particle
density, and velocity. Had we adjusted the jet parameters, such as the optical
depth and the width of the jet separately in each group, the agreement in
Fig.~\ref{bestmod} would have been even better. 

\begin{figure}
\centering
\includegraphics[angle=0,width=8cm]{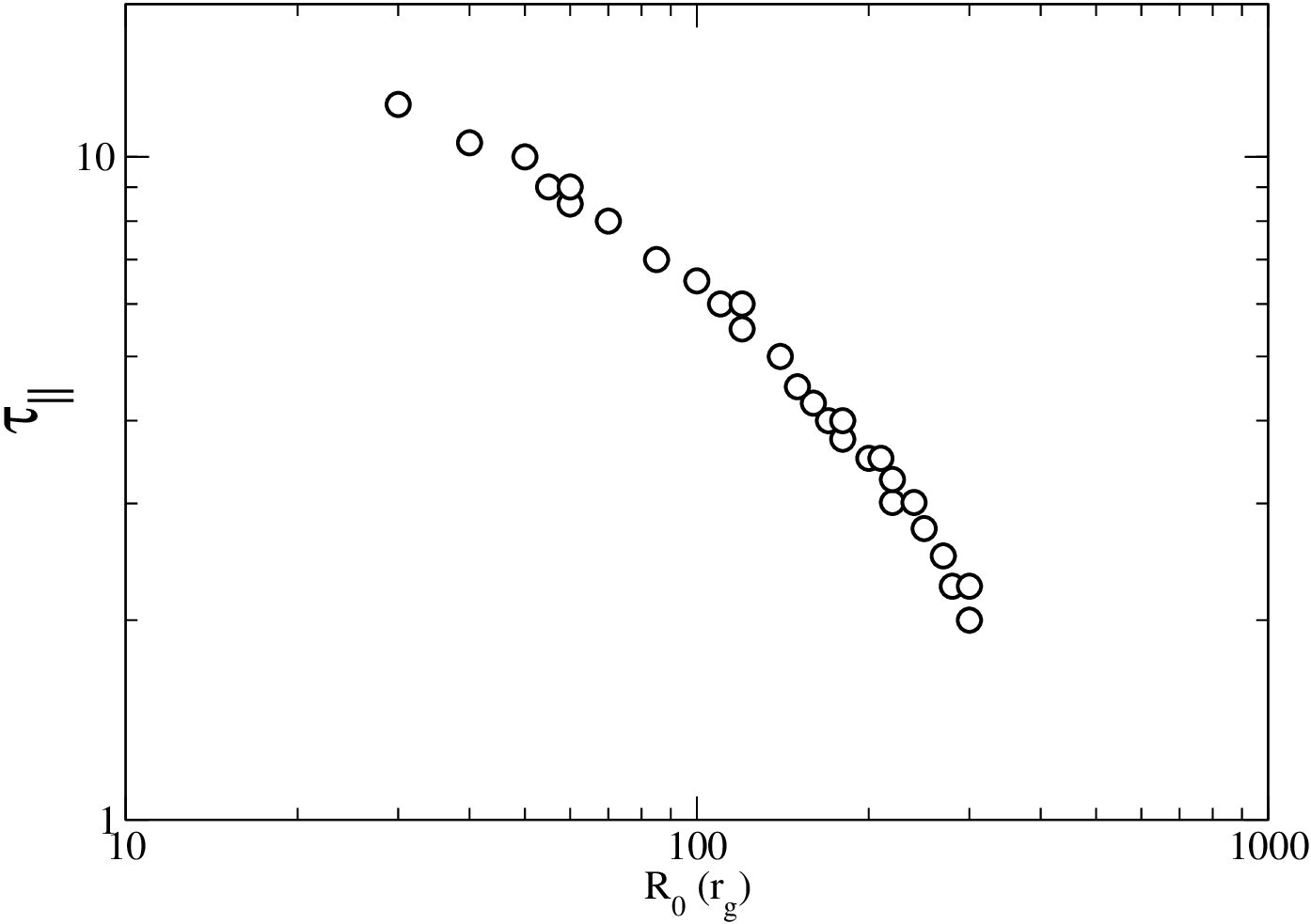} 
\caption{Relationship between the optical depth, $\tau_{\parallel}$ , and the 
width at the base of the jet, $R_0$ , for the models that reproduce the correlations. }
\label{tau-width}
\end{figure}

One very constraining relationship of our model is the fact that the parameters
$\tau_{\parallel}$ and $R_0$ of the best-fit models follow a tight correlation.
Although their actual absolute values may vary from source to source, the
optical depth and the jet width vary in  unison
\citep{kylafis08,reig18,kylafis18}. As the source moves up along the HS branch
(right branch in the HID), the Thomson optical depth along the jet decreases and
the width of the jet increases. The variation of the width of the jet with
luminosity is consistent with the idea that the accretion disk is truncated 
far away from the black hole during the HS and approaches it as the source
transits to a softer state.
   
Since the jet is fed by the hot inner flow, the width $R_0$ of the base of
the jet must be smaller than the extent of the hot inner flow, hence it must be
smaller than the distance $R_{\rm tr}$ of the inner part of the accretion disk
from the black hole.  At low luminosity in the HS, $R_0 << R_{\rm tr}$.  As the
luminosity increases in the HS, $R_0$ also increases, but without  exceeding
$R_{\rm tr}$.  At some luminosity, $R_0$ becomes comparable to $R_{\rm tr}$. 
This state would correspond to the transition between the HS and the HIMS. After
that, in the HIMS, and as the source moves horizontally to the left in this
branch (upper branch in the HID), the width decreases, while the optical depth
remains fairly constant or decreases slightly. The relationship between 
$\tau_{\parallel}$ and $R_0$ for the models used in this work is given in
Fig.~\ref{tau-width}.  Because we have restricted our analysis to observations
with ($\Gamma \simless 2$), the horizontal branch in the 
$\tau_{\parallel}$-$R_0$ plot  \citep[see Fig. 3 in][]{kylafis18} is absent in
Fig.~\ref{tau-width}.

Our jet model reproduces the expected trend between luminosity and truncation
radius and provides a clear physical and independent prediction on the disk
truncation radius. However, the values that we find at around the change of
state from HS into HIMS ($\sim 300\, r_g$) are somehow larger than those
obtained from spectral fits \citep{basak16,jiang19}, X-ray reverberation
\citep{demarco15} and QPOs \citep{ingram17}, which typically give $R_{\rm
tr}\simless 100\, r_g$. 

In closing, we stress that the correlations between spectral (photon index,
cutoff energy) and timing (time lag, rms variability, QPOs) parameters in BHBs
imply that they are coupled and strongly suggest that these components appear to
have a common underlying origin. The hysteresis of the HID adds another
constraint \citep{kylafis15}. So far,  only Comptonization in an extended jet
has been able to survive all these tests.

\section{Conclusion}

We have investigated the effect of the inclination on the correlation between
the time lag and the photon index in BHBs. We have shown that although the
correlation holds for all systems, there is a distinct inclination effect
affecting the relationship between the two variables. High-inclination systems
display, on average, a flatter correlation. This different behavior explains the
large scatter in the $t_{\rm lag}-\Gamma$ correlation reported by
\citet{reig18}, where no distinction for inclination was made.

We have simulated the process of inverse Compton 
scattering in a jet and generated
theoretical spectra and light curves. 
{\it The most remarkable result of this work is
the fact that we can reproduce the observed correlations between time lag and
photon index for systems with different inclination angles, with the same set of
models, by simply looking at the jet with different viewing angles.}

\begin{acknowledgements}
NDK thanks Asaf Pe'er for expressing his dislike for the large scatter 
exhibited in the time lag -- photon index correlation reported in
\citet{reig18}. The authors also thank I. Papadakis for fruitful discussions 
that helped improve the final version of this paper.
\end{acknowledgements}

\bibliographystyle{aa}
\bibliography{../../../bhb} 

\begin{thebibliography}{42}
\expandafter\ifx\csname natexlab\endcsname\relax\def\natexlab#1{#1}\fi

\bibitem[{{Akritas} \& {Bershady}(1996)}]{akritas96}
{Akritas}, M.~G. \& {Bershady}, M.~A. 1996, \apj, 470, 706

\bibitem[{{Bagri} {et~al.}(2018){Bagri}, {Misra}, {Rao}, {Singh Yadav}, \&
  {Pandey}}]{bagri18}
{Bagri}, K., {Misra}, R., {Rao}, A., {Singh Yadav}, J., \& {Pandey}, S.~K.
  2018, Research in Astronomy and Astrophysics, 18, 051

\bibitem[{{Band} \& {Grindlay}(1986)}]{band86}
{Band}, D.~L. \& {Grindlay}, J.~E. 1986, \apj, 311, 595

\bibitem[{{Basak} \& {Zdziarski}(2016)}]{basak16}
{Basak}, R. \& {Zdziarski}, A.~A. 2016, \mnras, 458, 2199

\bibitem[{{Belloni}(2010)}]{belloni10}
{Belloni}, T.~M. 2010, in Lecture Notes in Physics, Berlin Springer Verlag,
  Vol. 794, Lecture Notes in Physics, Berlin Springer Verlag, ed. T.~{Belloni},
  53

\bibitem[{{Cadolle Bel} {et~al.}(2007){Cadolle Bel}, {Rib{\'o}}, {Rodriguez},
  {Chaty}, {Corbel}, {Goldwurm}, {Frontera}, {Farinelli}, {D'Avanzo}, {Tarana},
  {Ubertini}, {Laurent}, {Goldoni}, \& {Mirabel}}]{cadolle07}
{Cadolle Bel}, M., {Rib{\'o}}, M., {Rodriguez}, J., {et~al.} 2007, \apj, 659,
  549

\bibitem[{{Chun} {et~al.}(2013){Chun}, {Din{\c c}er}, {Kalemci}, {G{\"u}ver},
  {Tomsick}, {Buxton}, {Brocksopp}, {Corbel}, \& {Cabrera-Lavers}}]{chun13}
{Chun}, Y.~Y., {Din{\c c}er}, T., {Kalemci}, E., {et~al.} 2013, \apj, 770, 10

\bibitem[{{De Marco} {et~al.}(2015){De Marco}, {Ponti}, {Mu{\~n}oz-Darias}, \&
  {Nandra}}]{demarco15}
{De Marco}, B., {Ponti}, G., {Mu{\~n}oz-Darias}, T., \& {Nandra}, K. 2015,
  \apj, 814, 50

\bibitem[{{Dunn} {et~al.}(2010){Dunn}, {Fender}, {K{\"o}rding}, {Belloni}, \&
  {Cabanac}}]{dunn10}
{Dunn}, R.~J.~H., {Fender}, R.~P., {K{\"o}rding}, E.~G., {Belloni}, T., \&
  {Cabanac}, C. 2010, \mnras, 403, 61

\bibitem[{{Esin} {et~al.}(1997){Esin}, {McClintock}, \& {Narayan}}]{esin97}
{Esin}, A.~A., {McClintock}, J.~E., \& {Narayan}, R. 1997, \apj, 489, 865

\bibitem[{{Garain} {et~al.}(2012){Garain}, {Ghosh}, \&
  {Chakrabarti}}]{garain12}
{Garain}, S.~K., {Ghosh}, H., \& {Chakrabarti}, S.~K. 2012, \apj, 758, 114

\bibitem[{{Georganopoulos} {et~al.}(2002){Georganopoulos}, {Aharonian}, \&
  {Kirk}}]{georganopoulos02}
{Georganopoulos}, M., {Aharonian}, F.~A., \& {Kirk}, J.~G. 2002, \aap, 388, L25

\bibitem[{{Ghosh} {et~al.}(2011){Ghosh}, {Garain}, {Giri}, \&
  {Chakrabarti}}]{ghosh11}
{Ghosh}, H., {Garain}, S.~K., {Giri}, K., \& {Chakrabarti}, S.~K. 2011, \mnras,
  416, 959

\bibitem[{{Giannios}(2005)}]{giannios05}
{Giannios}, D. 2005, \aap, 437, 1007

\bibitem[{{Hanke} {et~al.}(2009){Hanke}, {Wilms}, {Nowak}, {Pottschmidt},
  {Schulz}, \& {Lee}}]{hanke09}
{Hanke}, M., {Wilms}, J., {Nowak}, M.~A., {et~al.} 2009, \apj, 690, 330

\bibitem[{{Heil} {et~al.}(2015){Heil}, {Uttley}, \& {Klein-Wolt}}]{heil15}
{Heil}, L.~M., {Uttley}, P., \& {Klein-Wolt}, M. 2015, \mnras, 448, 3348

\bibitem[{{Homan} \& {Belloni}(2005)}]{homan05}
{Homan}, J. \& {Belloni}, T. 2005, \apss, 300, 107

\bibitem[{{Hua} \& {Titarchuk}(1995)}]{hua95}
{Hua}, X.-M. \& {Titarchuk}, L. 1995, \apj, 449, 188

\bibitem[{{Ingram} {et~al.}(2017){Ingram}, {van der Klis}, {Middleton},
  {Altamirano}, \& {Uttley}}]{ingram17}
{Ingram}, A., {van der Klis}, M., {Middleton}, M., {Altamirano}, D., \&
  {Uttley}, P. 2017, \mnras, 464, 2979

\bibitem[{{Jiang} {et~al.}(2019){Jiang}, {Fabian}, {Wang}, {Walton},
  {Garc{\'{\i}}a}, {Parker}, {Steiner}, \& {Tomsick}}]{jiang19}
{Jiang}, J., {Fabian}, A.~C., {Wang}, J., {et~al.} 2019, \mnras, 484, 1972

\bibitem[{{Jonker} {et~al.}(2012){Jonker}, {Miller-Jones}, {Homan}, {Tomsick},
  {Fender}, {Kaaret}, {Markoff}, \& {Gallo}}]{jonker12}
{Jonker}, P.~G., {Miller-Jones}, J.~C.~A., {Homan}, J., {et~al.} 2012, \mnras,
  423, 3308

\bibitem[{{Kuulkers} {et~al.}(2013){Kuulkers}, {Kouveliotou}, {Belloni},
  {Cadolle Bel}, {Chenevez}, {D{\'{\i}}az Trigo}, {Homan}, {Ibarra}, {Kennea},
  {Mu{\~n}oz-Darias}, {Ness}, {Parmar}, {Pollock}, {van den Heuvel}, \& {van
  der Horst}}]{kuulkers13}
{Kuulkers}, E., {Kouveliotou}, C., {Belloni}, T., {et~al.} 2013, \aap, 552, A32

\bibitem[{{Kylafis} \& {Belloni}(2015)}]{kylafis15}
{Kylafis}, N.~D. \& {Belloni}, T.~M. 2015, \aap, 574, A133

\bibitem[{{Kylafis} {et~al.}(2008){Kylafis}, {Papadakis}, {Reig}, {Giannios},
  \& {Pooley}}]{kylafis08}
{Kylafis}, N.~D., {Papadakis}, I.~E., {Reig}, P., {Giannios}, D., \& {Pooley},
  G.~G. 2008, \aap, 489, 481

\bibitem[{{Kylafis} \& {Reig}(2018)}]{kylafis18}
{Kylafis}, N.~D. \& {Reig}, P. 2018, \aap, 614, L5

\bibitem[{{Markoff} {et~al.}(2005){Markoff}, {Nowak}, \& {Wilms}}]{markoff05}
{Markoff}, S., {Nowak}, M.~A., \& {Wilms}, J. 2005, \apj, 635, 1203

\bibitem[{{Motta} {et~al.}(2018){Motta}, {Casella}, \& {Fender}}]{motta18}
{Motta}, S.~E., {Casella}, P., \& {Fender}, R.~P. 2018, \mnras, 478, 5159

\bibitem[{{Motta} {et~al.}(2015){Motta}, {Casella}, {Henze},
  {Mu{\~n}oz-Darias}, {Sanna}, {Fender}, \& {Belloni}}]{motta15}
{Motta}, S.~E., {Casella}, P., {Henze}, M., {et~al.} 2015, \mnras, 447, 2059

\bibitem[{{Mu{\~n}oz-Darias} {et~al.}(2013){Mu{\~n}oz-Darias}, {Coriat},
  {Plant}, {Ponti}, {Fender}, \& {Dunn}}]{munoz-darias13}
{Mu{\~n}oz-Darias}, T., {Coriat}, M., {Plant}, D.~S., {et~al.} 2013, \mnras,
  432, 1330

\bibitem[{{Narayan} \& {Yi}(1994)}]{narayan94}
{Narayan}, R. \& {Yi}, I. 1994, \apjl, 428, L13

\bibitem[{{Plant} {et~al.}(2014){Plant}, {Fender}, {Ponti}, {Mu{\~n}oz-Darias},
  \& {Coriat}}]{plant14}
{Plant}, D.~S., {Fender}, R.~P., {Ponti}, G., {Mu{\~n}oz-Darias}, T., \&
  {Coriat}, M. 2014, \mnras, 442, 1767

\bibitem[{{Ponti} {et~al.}(2012){Ponti}, {Fender}, {Begelman}, {Dunn},
  {Neilsen}, \& {Coriat}}]{ponti12}
{Ponti}, G., {Fender}, R.~P., {Begelman}, M.~C., {et~al.} 2012, \mnras, 422,
  L11

\bibitem[{{Reig} \& {Kylafis}(2015)}]{reig15}
{Reig}, P. \& {Kylafis}, N.~D. 2015, \aap, 584, A109

\bibitem[{{Reig} {et~al.}(2003){Reig}, {Kylafis}, \& {Giannios}}]{reig03}
{Reig}, P., {Kylafis}, N.~D., \& {Giannios}, D. 2003, \aap, 403, L15

\bibitem[{{Reig} {et~al.}(2018){Reig}, {Kylafis}, {Papadakis}, \&
  {Costado}}]{reig18}
{Reig}, P., {Kylafis}, N.~D., {Papadakis}, I.~E., \& {Costado}, M.~T. 2018,
  \mnras, 473, 4644

\bibitem[{{Remillard} \& {McClintock}(2006)}]{remillard06}
{Remillard}, R.~A. \& {McClintock}, J.~E. 2006, \araa, 44, 49

\bibitem[{{Russell} {et~al.}(2014){Russell}, {Soria}, {Miller-Jones}, {Curran},
  {Markoff}, {Russell}, \& {Sivakoff}}]{russell14}
{Russell}, T.~D., {Soria}, R., {Miller-Jones}, J.~C.~A., {et~al.} 2014, \mnras,
  439, 1390

\bibitem[{{Stiele} {et~al.}(2012){Stiele}, {Mu{\~n}oz-Darias}, {Motta}, \&
  {Belloni}}]{stiele12}
{Stiele}, H., {Mu{\~n}oz-Darias}, T., {Motta}, S., \& {Belloni}, T.~M. 2012,
  \mnras, 422, 679

\bibitem[{{Sunyaev} \& {Titarchuk}(1980)}]{titarchuk80}
{Sunyaev}, R.~A. \& {Titarchuk}, L.~G. 1980, \aap, 86, 121

\bibitem[{{Tetarenko} {et~al.}(2016){Tetarenko}, {Sivakoff}, {Heinke}, \&
  {Gladstone}}]{tetarenko16}
{Tetarenko}, B.~E., {Sivakoff}, G.~R., {Heinke}, C.~O., \& {Gladstone}, J.~C.
  2016, \apjs, 222, 15

\bibitem[{{van den Eijnden} {et~al.}(2017){van den Eijnden}, {Ingram},
  {Uttley}, {Motta}, {Belloni}, \& {Gardenier}}]{vandereijnden17}
{van den Eijnden}, J., {Ingram}, A., {Uttley}, P., {et~al.} 2017, \mnras, 464,
  2643

\bibitem[{{Zdziarski}(1998)}]{zdziarski98}
{Zdziarski}, A.~A. 1998, \mnras, 296, L51

\end{thebibliography}

\end{document}